\documentclass[prb,twocolumn,showpacs,preprintnumbers,amsmath,amssymb]{revtex4}

\usepackage{graphicx}
\usepackage{dcolumn}
 
\newcommand{\beq}{\begin{equation}} 
\newcommand{\eeq}{\end{equation}} 
\newcommand{\beqa}{\begin{eqnarray}} 
\newcommand{\eeqa}{\end{eqnarray}}

\begin{document} 
\title{Pairing and Superconductivity from weak to strong coupling in the 
Attractive Hubbard model}
\author{A. Toschi$^1$, P. Barone$^{1,2}$, M. Capone$^{3,1,4}$, and C.
 Castellani$^1$}  
\affiliation{$^1$ 
INFM Center for Statistical Mechanics and Complexity SMC and  
Dipartimento di Fisica, Universit\`a di Roma ``La Sapienza'',
Piazzale Aldo Moro 2, I-00185 Roma, Italy}
\affiliation{$^2$Dipartimento di Fisica, Universit\a di Roma 3, Via della 
Vasca Navale 84,
I-00146, Roma, Italy}
\affiliation{$^3$Istituto dei Sistemi Complessi (ISC) del CNR, Via dei Taurini 19,
00185, Roma, Italy} 
\affiliation{$^4$Enrico Fermi Center, Roma, Italy} 

\begin{abstract}
The finite-temperature phase diagram of the 
attractive Hubbard model is studied by means of the Dynamical
Mean Field Theory. 
We first consider the normal phase of the model by explicitly frustrating 
the superconducting ordering. In this case we obtain a first-order
pairing transition between a metallic phase and a paired phase
formed by strongly coupled incoherent pairs.
The transition line ends in a finite temperature critical point, but a
crossover between two qualitatively different 
solutions still occurs at higher temperature.
Comparing the superconducting and the normal phase solutions, we find that 
the superconducting instability always occurs before the pairing transition 
in the normal phase takes place, i.e., $T_c > T_{pairing}$. Nevertheless, 
the high-temperature phase diagram at $T > T_c$ is still characterized 
by a crossover from a metallic phase to a preformed pair phase. 
We characterize this crossover by computing 
different observables that can be used to identify the pseudogap region, like
the spin susceptibility, the specific heat and the single-particle spectral 
function.

\end{abstract}

\date{\today} 
\maketitle 

\section{Introduction}

The attractive Hubbard model represents an unvaluable tool to 
understand properties of pairing and superconductivity in systems
with attractive interactions. The simplifications introduced in this model
allow a comprehensive study of the evolution from the weak-coupling 
regime, where superconductivity is due to BCS pairing in
a Fermi liquid phase, and a strong coupling regime, in which the system
is better described in terms of bosonic pairs, whose condensation
gives rise to superconductivity (Bose Einstein (BE) superconductivity)
\cite{micnas}.
It has been convincingly shown that such an evolution is a smooth crossover
and the highest critical temperature is achieved in the intermediate
regime where none of the limiting approaches is rigorously 
valid\cite{micnas,bcsbeqmc}.
A  realization of such a crossover scenario has been
recently obtained through the development of experiments on the condensation
of ultracold trapped fermionic atoms\cite{atomi}. 
In these systems the strength of the
attraction can be tuned by means of a tunable Fano-Feschbach
resonance, and the whole crossover can be described\cite{xoveratomi}.

In the context of high-temperature superconductivity, 
the intermediate-strong coupling regime in which incoherent pairs 
are formed well above the critical temperature has been invoked as an 
interpretation of the pseudogap phase\cite{bcsbeqmc}. Moreover, since 
the early days of the discovery of these materials, 
the evolution with the doping level of both the normal- and the 
superconducting-phase properties  
induced some authors\cite{bcsbemix,bcsbestrin} to recognize the fingerprints 
of a crossover between a relatively standard BCS-like superconductivity in 
the overdoped materials and a strong-coupling superconductivity associated
to Bose-Einstein condensation (BE) in the underdoped materials.
Indeed at optimal doping the zero-temperature coherence
 length is estimated to be around 
10$-$20 \AA \cite{pan,iguchi}, i.e., much smaller than 
for conventional superconductors but  still large enough to 
exclude the formation of local pairs.\cite{crossover,uemura}

It is understood that the attractive Hubbard model has not to be taken as
a microscopic model for the cuprates, since
a realistic description of the copper-oxygen planes of these materials
unavoidably requires a proper treatment of strong Coulomb repulsion.
This simplified model represents instead an ideal framework where the
evolution from weak to strong coupling can be studied 
by simply tuning the strength of the attraction. 
The main aim of the present work is to identify if, and to which extent,
at least some aspects of the phenomenology of the cuprates can be interpreted 
simply in terms of a crossover from weak to strong coupling.

The main simplifications introduced by the attractive Hubbard model
 can be summarized as {\it (i)} Neglect of repulsion. Even if some attraction 
has to develop at
low energy, the large short-range Coulomb repulsion implies that
the interaction must become repulsive at high-energy in real systems. 
In some sense, 
an attractive Hubbard model picture can at most be applied to the
low-energy quasiparticles.
{\it{(ii)}} The model naturally presents s-wave superconductivity, as opposed
to the d-wave symmetry observed in the cuprates
{\it{(iii)}} Neglect of retardation effects. The Hubbard model describes
instantaneous interactions, while every physical pairing is expected to
present a typical energy scale.

The model is written as
\begin{eqnarray}
\label{hubbard}
{\cal H} &=& -t \sum_{<ij>\sigma} c_{i\sigma}^{\dagger} c_{j\sigma} 
-U\sum_{i}\left ( n_{i\uparrow}-\frac{1}{2}\right )
\left ( n_{i\downarrow}-\frac{1}{2}\right )+\nonumber\\
& & -\mu\sum_i (n_{i\uparrow}+n_{i\downarrow})
\end{eqnarray} 
where $c_{i\sigma}^{\dagger}$ ($c_{i\sigma}$) creates (destroys) 
an electron with spin $\sigma$ on the site $i$ and $n_{i\sigma} = 
c_{i\sigma}^{\dagger}c_{i\sigma}$ is the number operator; 
$t$ is the hopping amplitude and $U$ is the Hubbard on-site attraction
(we take $U > 0$, with an explicit  minus sign in the hamiltonian). 
Notice that, with this notations, the Hamiltonian is explicitly
particle-hole symmetric for  $\mu = 0$, which therefore corresponds to 
$n=1$ (half-filling). 

Despite its formal simplicity, this model can be solved exactly only in 
$d = 1$, while in larger dimensionality analytical calculations
are typically limited to weak ($U \ll t$)
or strong ($U \gg t$) coupling, where the BCS and the BE approaches 
are reliable approximations.
It is anyway known that for $d \ge 1$, 
the ground state of (\ref{hubbard})
is superconducting for all values of $U$ and all densities $n$, with the
only exception of the one-dimensional half-filled case.   
At half-filling the model has an extra-symmetry and the superconducting 
and the  charge-density-wave order parameters become degenerate.

A reliable description of the evolution of the physics as a function of 
$U$ requires to treat the two limiting regimes on equal footing
overcoming the drawbacks of perturbative expansions. 
Quantum Monte Carlo (QMC) simulations represent a valuable tool in this regard,
and they have been applied to the two\cite{bcsbeqmc,moreo,singerold,
singernew,angleres} and 
three\cite{sewer} dimensional attractive Hubbard
model. Even if the sign problem does not affect these simulations, finite 
size effects and memory requirements still partially limit the potentiality 
of this approach.

A different non perturbative approach is the  Dynamical Mean-Field Theory 
(DMFT), 
that neglects the spatial correlations beyond the
mean field level in order to  fully retain the local quantum 
dynamics, and becomes exact in the limit of infinite dimensions\cite{dmft}. 
Due to the local nature of the interaction in the attractive Hubbard model,
we expect that the physics of local pairing is well described in DMFT.
Moreover, this approach is not biased toward metallic or insulating states,
and it is therefore particularly useful to analyze the BE-BCS crossover.
On the other hand, the simplifications introduced by the DMFT are 
rigorously valid only in the infinite dimensionality limit, and even if 
the DMFT has obtained many successes for three dimensional systems,
its relevance to lower dimensionality like $d=2$ is much less 
established, and represents a fourth limitation of our study in light
of a comparison with the physics of the cuprates.
In particular, the role of dimensionality in determining
the pseudogap properties
of the attractive Hubbard model has been discussed in Refs. 
\cite{dimensionalita}.

The study of the attractive Hubbard model can greatly benefit 
of a mapping onto a repulsive model in a magnetic field. 
The mapping is realized in a bipartite 
lattice\cite{auerbach} by a 'staggered' particle-hole transformation on 
the down spins $c_{i\downarrow} \to (-1)^i {c}_{i\downarrow}^\dagger$.
The attractive model with 
a finite density $n$ transforms into a half-filled repulsive model with a 
finite magnetization $m = n-1$. The 
chemical potential is transformed, accordingly, into a 
 magnetic field $h = \mu$. 
In the $n=1$ case (half-filling) the two models are therefore
completely equivalent. We notice that the above mapping does not
only hold for the normal phases, but extends to the broken symmetry 
solutions. The three components of the antiferromagnetic order
parameter of the repulsive Hubbard model are in fact mapped onto 
a staggered charge-density-wave parameter ($z$ component of the spin)
and an s-wave superconducting order parameter ($x-y$ components).
The above mapping is extremely useful, since it allows to exploit
all the known results for the
repulsive model and for the Mott-Hubbard transition 
to improve our understanding of the attractive model.

In recent works the DMFT has been used to study the normal phases
of the attractive Hubbard model. In particular, a phase transition
has been found both at finite\cite{keller} and at zero temperature\cite{prl}
between a metallic solution and an pairing phase of pairs.
The insulating pairs phase is nothing but a realization of a superconductor
without phase coherence, i.e., a collection of independent pairs.
As it has been discussed in Ref.\cite{prl,keller}, this phase is the
'negative-$U$' counterpart of the paramagnetic Mott insulator found for the
repulsive Hubbard model.
We notice that the insulating character of the pairing phase is a 
limitation of the DMFT approach, in which the residual kinetic energy
of the preformed pairs is not described.
The pairing transition has been first identified in Ref. \cite{keller}
by means of a finite temperature QMC solution of the 
DMFT.
The $T=0$ study of Ref. \cite{prl} has clarified that the 
pairing transition is always of first order except for the half-filled
case, and that it takes
place with a finite value of the 
quasiparticle weight $Z = (1-\partial \Sigma(\omega)/\partial \omega)^{-1}$,
associated to a finite spectral weight at the Fermi level.
In the latter paper, it has also been shown that the pairing transition
gives rise to phase separation.

For what concerns the onset of superconductivity, a DMFT calculation of the 
critical temperature $T_c$ has been 
performed for the case of $n=0.5$ in the same Ref. \cite{keller}. 
The $T_c$ curve, extracted from the divergence of the pair-correlation 
function 
in the normal phase, displays a clear maximum at intermediate coupling and
reproduces correctly both the BCS and the BE predictions in 
the asymptotic limits, remaining finite for all $U \neq 0$. 

In this work we complement the analysis of Ref. \cite{prl}, by extending
our phase diagram to finite temperature, still using Exact
Diagonalization (ED) to solve the impurity model associated with 
the DMFT of the Hubbard model\cite{caffarel}. 
We also compare the normal state solutions with the 
superconducting solutions which are stable at low temperatures. 
The use of ED allows us to reach arbitrarily small temperatures which
are hardly accessible by means of QMC. Quite naturally,
the extension of ED to finite temperature requires a more severe
truncation of the Hilbert space. We have checked that all the 
thermodynamical quantities we show are only weakly dependent on the
truncation. 
The plan of the paper is the following: in Sec. II we briefly introduce the 
DMFT method and its generalization to the superconducting phase;
In Sec. III we discuss the finite temperature phase diagram in the normal phase
characterizing the low-temperature pairing transition; In Sec. IV we analyze
the superconducting solutions; In Sec. V we compare different estimators
of the pseudogap temperature in the high-temperature normal phase. Sec. VI 
contains our concluding remarks.

\section{method}

The DMFT extends the concept of classical mean-field theories to quantum
problems, by describing a lattice model in terms of an effective
dynamical local theory.
The latter can be represented through an impurity model subject
to a self-consistency condition, which contains all the information
about the original lattice structure through the non-interacting 
density of states (DOS)\cite{dmft}.
Starting from the Hubbard model (\ref{hubbard}), we obtain an attractive
Anderson impurity model
\begin{eqnarray}
\label{aim}
{\cal H}_{AM} &=& -\sum_{k,\sigma} V_k c^{\dagger}_{k,\sigma} c_{0,\sigma} +
  H.c. + \sum_{k,\sigma} \epsilon_k c^{\dagger}_{k,\sigma} c_{k,\sigma} 
\nonumber\\ 
&-& U\left ( n_{0\uparrow}-\frac{1}{2}\right )
\left ( n_{0\downarrow}-\frac{1}{2}\right )+\mu n_0,
\end{eqnarray}
The self-consistency is expressed by requiring the identity between
the local self-energy of the lattice model and the impurity self-energy
\begin{equation}
\label{sigma}
\Sigma(i\omega_n) = {\cal G}^0(i\omega_n)^{-1} - G(i\omega_n)^{-1},
\end{equation}
where $G(i\omega_n)$ is the local Green's function of (\ref{aim}), and
${\cal G}^0(i\omega_n)^{-1}$ is the dynamical Weiss field, related to the
parameters in (\ref{aim}) by
\begin{equation}
\label{g0}
{\cal G}^0(i\omega_n)^{-1}= i\omega_n +\mu -\sum_k\frac{V_k^2}{i\omega_n - 
\epsilon_k}.
\end{equation}
By expressing the local component of the Green's function in terms of the
lattice Green's function, namely  $G(r=0,i\omega_n)= \sum_k G(k,i\omega_n)$, 
Eq. (\ref{sigma}) implies
\begin{equation}
\label{selfconsistence}
{\cal G}^0(i\omega_n)^{-1} = \left ( \int d\epsilon 
\frac{D(\epsilon)}{i\omega_n + \mu -\epsilon - \Sigma(i\omega_n)} 
\right )^{-1} + \Sigma(i\omega_n),
\end{equation}
where $D(\epsilon)$ is the non interacting density of states of the original 
lattice.
We consider the infinite-coordination Bethe lattice, 
with semicircular DOS of half-bandwidth $D$ (i.e.,
$D(\epsilon) = (2/\pi D^2) \sqrt{D^2 - \epsilon^2}$), for which 
Eq. (\ref{selfconsistence}) 
is greatly simplified and becomes
\begin{equation}
\label{selfbethe}
{\cal G}^0(i\omega_n)^{-1} = i\omega_n + \mu - \frac{D^2}{4}G(i\omega_n).
\end{equation}

In this work we also consider solutions with explicit s-wave superconducting
order, by allowing for local anomalous Green's functions
$F(\tau)  = -\langle T_{\tau} c_{0\uparrow}(\tau) c_{0\downarrow}\rangle$.
The whole DMFT formalism can then be recast in Nambu-Gor\'kov spinorial
representation\cite{dmft}, and Eqs. (\ref{sigma}) and (\ref{selfconsistence}) 
must be read as matrix identities in the Nambu space. 
As far as the impurity model is concerned, we need to describe an
Anderson impurity model with a superconducting bath or, equivalently, 
with an anomalous hybridization in which Cooper pairs are created and
destroyed in the electronic bath, i.e., a term 
$\sum_k V^s_k (c_{k\uparrow}c_{k\downarrow} + H.c.)$ is added to (\ref{aim}).

The heaviest step of the DMFT approach is to compute $G(i\omega_n)$ 
for the Anderson
model (\ref{aim}). This solution requires either a numerical approach or
some approximation. Here we use Exact Diagonalization.
Namely, we discretize the Anderson model, by truncating the sums over $k$
in Eqs. (\ref{aim}) and (\ref{g0}) to a finite number of levels
$N_s$. It has been shown that extremely small values of $N_s$ provide 
really good results for thermodynamic properties and reliable results
for spectral functions.
In this work we use the ED approach at finite temperature, where it is
not possible to use the Lanczos algorithm, which allows to find the
groundstate of extremely large matrices. To obtain the full spectrum
of the Hamiltonian, needed to compute the finite-temperature properties,
we are forced to a rather small value of $N_s$, up to 6. 
All the results presented here are for $N_s = 6$, and
we always checked that changing $N_s$ from 5 to 6 does not affect the
relevant observables we discuss in the present work, except for the
real-frequency spectral properties.

\section{The pairing Transition} 
In this section we limit our analysis to normal phase paramagnetic solutions 
in which no superconducting ordering is allowed. 
Even if the s-wave superconducting solution is expected to be the stable one 
at low temperatures, our normal state
solutions are representative of the normal phase above the critical 
temperature.
The region in which the normal state is stable may of course be enlarged by
frustrating superconductivity through, e.g., a magnetic field. 
Moreover, the nature of the normal phase gives important indications 
on the nature of the pairing in the different regions of the phase diagram.
As mentioned above, it has been shown that the normal phase of the 
attractive Hubbard model is characterized by a ``pairing'' transition
between a Fermi-liquid phase and a phase in which the electrons are paired,
but without any phase coherence among the pairs.

The pairing transition has been first discussed at finite temperature
in Ref. \cite{keller}, and a complete characterization at $T=0$ has been
given in Ref. \cite{prl}.
In this paper we complete the finite temperature study of the 
transition and connect it to the zero-temperature phase diagram,
finally drawing a complete phase diagram in the attraction-temperature
plane for a density  $n=0.75$, taken as representative
of a generic density (except for the peculiar particle-hole
symmetric $n=1$ case). This situation would 
correspond to a repulsive model at half-filling in an external magnetic 
field tuned to give a finite magnetization $m=0.25$.
The $T=0$ DMFT solution of the attractive Hubbard model is characterized by 
the existence of two distinct solutions, a metallic one with
a finite spectral weight at the Fermi level and an insulating solution
formed by pairs, with no weight at the Fermi level.
The previous study has also clarified that the quasiparticle weight
$Z = (1-\partial\Sigma(\omega)/\partial\omega)^{-1}$, which may be used
as a sort of order parameter for the Mott transition at half-filling,
loses this role for the doped attractive Hubbard model, being it finite 
both in the metallic and pairing phases. 
At $T=0$, the metallic solution exists only for $U < U_{c2}$, and
the insulating one for $U > U_{c1}$, with $U_{c1}  < U_{c2}$. In other
words, a coexistence region is present where both solutions exist, and
where the actual ground state is determined minimizing the 
internal energy.
\begin{figure}[htbp]
\begin{center}
\includegraphics[width=6.5cm]{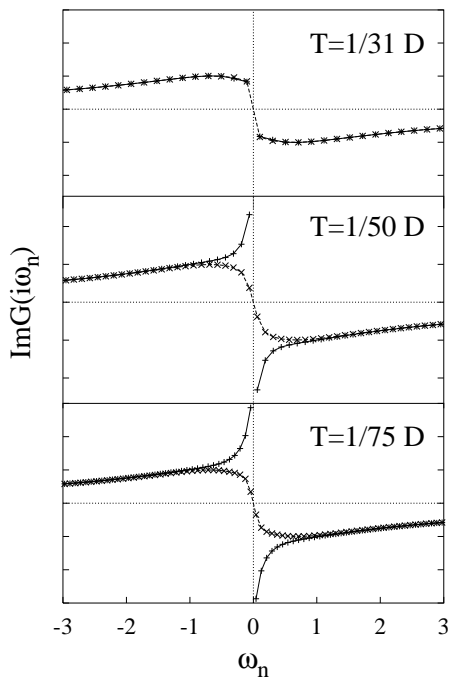}
\end{center}
\caption{Evolution of the imaginary part of the Green's function as a 
function of temperature for $U/D = 2.4$. 
In each panel are shown the metallic ($+$) and insulating ($\times$) solutions.
(the chosen value of the attraction lies 
in the coexistence region).}
\label{img}
\end{figure}
\begin{figure}[htbp]
\begin{center}
\includegraphics[width=6.5cm]{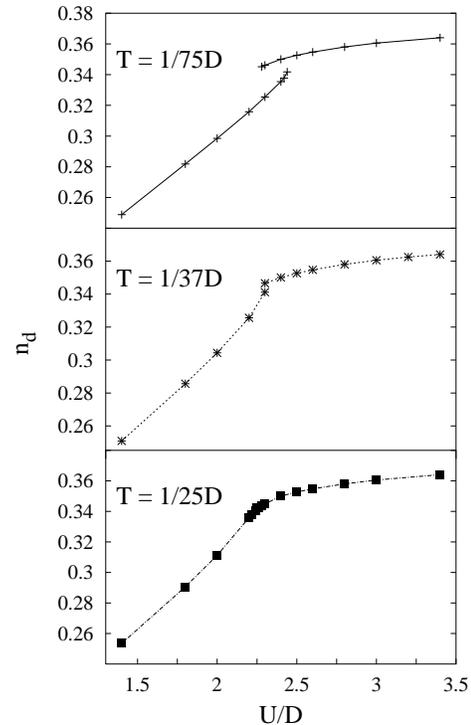}
\end{center}
\caption{Average double occupation as a function of the attraction strength 
for different temperatures. The first order transition at low temperatures
becomes a continuous evolution at high temperatures, where there is no more
distinction between metallic and pairing solutions.}
\label{doubocc}
\end{figure}
The clear-cut $T=0$ characterization of the two solutions based by the 
low-energy spectral weight
is lost at finite temperature, where both solution have finite weight at the 
Fermi level. Nonetheless, two families of solutions can still be defined, each 
family being obtained by continuous evolution of the different $T=0$ phases.
The two solutions are still clearly identified 
at relatively low temperatures, further increasing the 
temperature, the differences between the two solutions is gradually washed out,
as shown in Fig. \ref{img}, where we plot the temperature
evolution of the imaginary part of the Green's function in imaginary
frequency for $U= 2.4D$, which lies in the $T=0$ coexistence region.
While at $T=1/75D$ and $T=1/50D$ the difference in the two solutions is 
still clear, at $T=1/31D$, the two solutions become basically 
indistinguishable.
This result suggests that, as intuitively expected, the temperature 
reduces the difference between the solutions and consequently, the size of
the coexistence region, which is expected to close at some finite 
temperature critical point (the attractive counterpart of the 
endpoint of the line of metal-insulator transitions in the
repulsive model \cite{lange}).
A similar information is carried by the analysis of the average value of 
double occupancy $n_d = \langle n_{\uparrow} n_{\downarrow}\rangle$. 
This quantity naturally discriminates between an pairing phase with
a large value of  $n_d$ and a metal with a smaller value. 
As shown in Fig. \ref{doubocc}, at low temperature we have two solutions with 
a different value of $n_d$ in the coexistence region, and a jump in this 
quantity at the transition. Upon increasing the temperature, the two solutions
tend to join smoothly one onto the other, signaling again the closure 
of the coexistence region, which is substituted by a crossover region.
Analogous behavior is displayed by the quasiparticle weight $Z$.

\begin{figure}[htbp]
\begin{center}
\includegraphics[width=8.5cm]{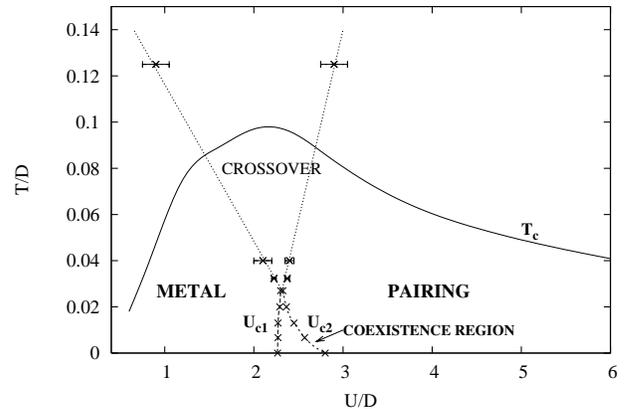}
\end{center}
\caption{Phase diagram in the $U$-$T$ plane. At low temperature two critical 
lines $U_{c1}(T)$ and $U_{c2}(T)$ individuate the coexistence region. 
The two lines converge in a finite temperature critical point. At higher 
temperatures we can still define two crossover lines. 
The superconducting critical temperature is also drawn as a solid line (cfr. 
Fig. 4).}
\label{phd}
\end{figure}

Repeating the same analysis for a wide range of coupling constants
and temperatures, 
we are able to construct a finite-temperature phase diagram for the 
pairing transition, shown in Fig. \ref{phd}.
For temperatures smaller than a critical temperature $T_{pairing}$, 
we compute the finite temperature extensions of $U_{c1}$ and $U_{c2}$,
which mark the boundary of the coexistence region.
The two lines (depicted as dashed lines in Fig. 3) converge into 
a finite temperature critical point at $U = U_{pairing} \simeq 2.3 D$ and 
$T=T_{pairing} \simeq 0.03 D $. 
Despite the closure of the coexistence
region, a qualitative difference between weak coupling and strong coupling
solutions can still be identified for $T > T_{pairing}$, 
determining a crossover region in which 
the character of the solution smoothly evolves from one limit to the other
as the attraction is tuned.
At this stage, the crossover region
is ``negatively'' defined as the range in which the Green's function does not
resemble any of the two low temperature phases. The crossover lines are 
estimated as the points in which it becomes impossible to infer from the 
Matsubara frequency Green's function whether the low-energy behavior is 
metallic or insulating. 
It has been shown for the repulsive Hubbard model that this kind of 
crossover is accompanied by a qualitative difference in transport 
properties. In the region on the left of the crossover, the conduction is
metallic and the resistivity increases with temperature. In the 
intermediate crossover region the system behaves like a semiconductor with a 
resistivity which decreases upon heating, and finally in the phase on the right
of the crossover region the system behaves like a heated insulator\cite{dmft}.

Coming from the left, the first
crossover occurs when $G(i\omega_n)$ has no longer a clear metallic behavior
with a finite value at zero frequency, while the second crossover line delimits
the region in which the gap of the paired solution is closed by thermal 
excitations. We will come back later to the crossover region and compare the
above defined lines with physically sensible estimators of the 
pseudogap temperature, like the specific heat and the spin susceptibility.

Turning to the coexistence region, we can also ask ourselves which 
is the stable phase.
This requires a comparison between the Gibbs free energies of the
two phases. At half-filling, where the attractive and the repulsive model
are equivalent, it has been shown that at $T=0$ the 
metallic solution is stable in the whole coexistence region\cite{moeller}.
At finite temperature  it has been shown numerically that
 the insulator becomes stable in a large
portion of the coexistence region due to its large entropy\cite{dmft}.
The transition is therefore of first order for all temperature below the
critical temperature, except for the two second-order endpoints 
at $T=0$ and $T=T_{pairing}$.
For densities out of half-filling it has been shown in Ref. \onlinecite{prl} 
that the transition  is of first order 
already at $T=0$ and it is accompanied by a small phase separation region. 
For $n=0.75$, the $T=0$ first-order transition occurs quite close to $U_{c2}$.

Analogously to the half-filling case, the finite temperature almost
immediately favors the pairing phase. Indeed, computing the free
energy following, e.g., Ref. \cite{gabilandau}, we find 
the pairing phase stable for almost every point in the coexistence region.
We had to use an extremely dense mesh of points in the $U$ direction
to identify a small section where the metallic phase is stable at finite 
temperature. 
Therefore the finite temperature first-order transition occurs 
extremely close to the $U_{c1}$ line for finite temperature and rapidly moves 
closer to $U_{c2}$ only at really small temperatures.

\section{The Superconducting Phase}

The above stability analysis has been restricted to normal phase solutions.
Indeed the superconducting solution is expected to be the stable one 
at $T=0$ for all  densities and values of the interaction $U$. 
The critical temperature $T_c$ is obtained directly as the highest temperature
for which a non-vanishing anomalous Green's function $F(\omega)$ exists. 

\begin{figure}[htbp]
\begin{center}
\includegraphics[width=8.5cm]{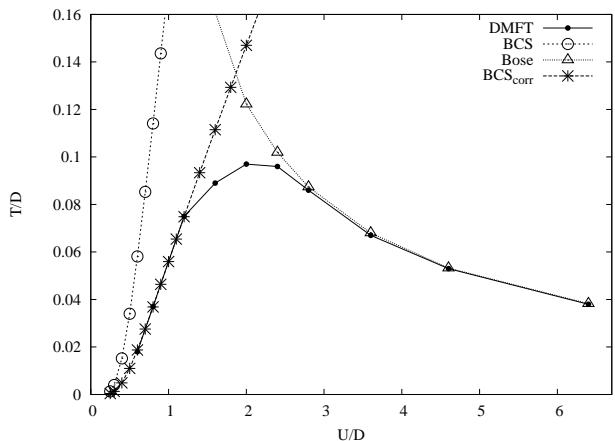}
\end{center}
\caption{Critical temperature as a function of $U$ at $n=0.75$: 
the DMFT data (black circle) are compared with the
 BCS -both the bare (empty circles) and the rescaled one (stars)- and the  
BE mean field predictions (empty triangles) for an hard-core boson 
systems (see Refs. \onlinecite{micnas,kellerlow}).}
\label{tc}
\end{figure}

The DMFT critical temperature $T_c$ for $n=0.75$ as a function of $U$ is 
reported in 
Fig. \ref{tc} (full dots) and it qualitatively 
reproduces the limiting behavior,
with an exponential BCS-like behavior for small U's and a $1/U$ decrease at
 large
$U$ according to the  expression 
for the BE condensation temperature  hard-core boson system\cite{kellerlow}.
As a result, $T_c$ assumes its maximum value of about $0.1 D$ for an
 intermediate coupling strength $U_{max} \simeq 2.1 D$. 
Interestingly, the maximum $T_c$ occurs
almost exactly at the coupling for which the pairing transition in the
normal phase would take place in the absence of superconductivity. 

It might be noticed however that, while the BE result (open triangles in Fig. 
\ref{tc}) basically falls on top of the DMFT results, the BCS formula 
(open circles) only qualitatively follows the full solution. This
``asymmetry'' in recovering the BCS behavior 
arises from the partial screening of the bare attraction due to 
second order polarization terms\cite{viverit}. 
Because of these corrections  the attraction is renormalized as
$U_{eff} \simeq U - A U^2/t$, so that $\frac{1}{U_{eff}} \simeq \frac{1}
{U}(1+AU/t) = 
1/U +A/t$.
When this correction is plugged in the BCS formula for $T_c$, it results
in a correction to the prefactor. 
If we simply extract the rescaling factor for a given small  value of $U$
($T_c/T_c^{BCS} \simeq 0.32$) and we simply scale the whole weak-coupling
 curve by this 
factor, we obtain the points marked with asterisks, whose agreement 
with the DMFT results
does not require further comments. It is interesting, instead, to compare 
the DMFT estimations for $T_c$ with the QMC results: 
despite the presence of many factors (such as the exact shape of the D.O.S. 
of the model or the finite dimension effects) which are capable to introduce 
relevant variations in the  values of $T_c$, some general similarities appear 
clearly. Indeed, while simple rescaling the data in terms of 
the half-bandwidth $D$, both $T_c$ and $U_{max}$ estimations with the 
two\cite{singerold,singernew} and three\cite{sewer} dimensional QMC 
are lower than the DMFT evaluation (i.e., 
$T_c \sim 0.04 D $, $U_{max} \sim 0.7 D$ for the $d=2$ case, and 
$T_c  \sim 0.05 D$, $U_{max} \sim 1.3 D$ in $d=3$, 
even if for a lower density of $n=0.5$), one can observe, quite surprisingly,
that the ratio between $T_c$ and $U_{max}$ is around $0.04 \div 0.05$ in both 
the DMFT and the two QMC cases. 
    
Coming back to our DMFT results, the simplest and most important 
observation is that
the critical temperature is always higher than the critical temperature
for the pairing transition.
For example in our $n=0.75$ case,  
$T_c^{max}$ is about $ 0.1 D$, against a  $T_{pairing}$ of $0.03D$.
As a result, the whole phase transition is hidden by superconductivity,
which remains as the only real instability of the system (cfr. Fig. 3). 
Nevertheless, the crossover lines at higher temperature survive the 
onset of superconductivity. Therefore the normal phase we reach for  
$T > T_c$ is really different according to the regime of coupling we are in.
For weak-coupling, the normal phase is substantially a regular Fermi-liquid
and superconductivity occurs as the standard BCS instability. In the
strong-coupling regime, the normal phase is instead a more correlated phase
which presents a pseudogap in the spectrum. 
At intermediate coupling, where the superconducting critical temperature
reaches its maximum, the normal phase is in a crossover region between 
the two limiting behaviors.

\section{The Pseudogap Phase: Spin Susceptibility, Specific Heat and 
Spectral Functions}

Even if the onset of superconductivity completely hides the pairing 
transition, the fingerprints of the low-temperature normal phase
are still visible in the high-temperature phase diagram, in which
a crossover from a metallic phase to a gapped phase is still present.
It is tempting to associate the region in which the system behaves
as a collection of incoherent pairs to the pseudogap regime of the cuprates. 
It is important to underline that, in this framework, the definition of the 
pseudogap phase is somewhat tricky, and it implies a certain degree of
arbitrariness. In this section we come back to this region and compute 
various observable whose anomalies have been used to identify the 
pseudogap phase and compare the related estimates of the pseudogap
temperature $T^*$.

\begin{figure}[htbp]
\begin{center}
\includegraphics[width=8.5cm]{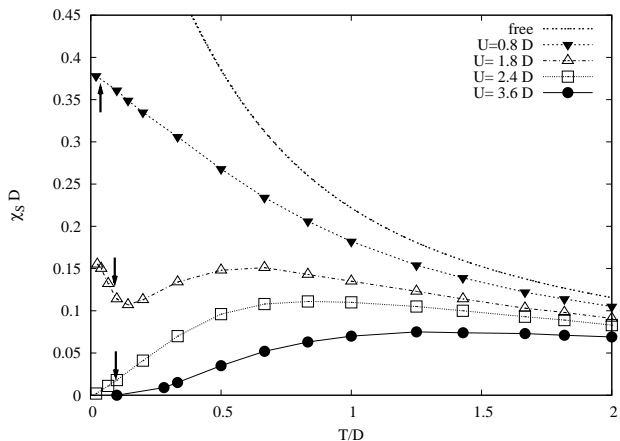}
\end{center}
\caption{Spin susceptibility in the normal phase as a function of 
temperature for $U/D= 0.0, 0.8,1.8,2.4,3.6$. 
These values of $U$ are representative of all
 the interesting region of the phase diagram in Fig. \ref{phd}, moving from 
the metallic to the paired side. The values of the 
superconducting critical temperature are marked by small black arrows.}
\label{chispin}
\end{figure}

Our first estimate is based on the evaluation of the uniform spin
susceptibility $\chi_s$
as a function of temperature for different attraction strengths.
The opening of a gap in the spin excitation spectrum, not associated with 
any long-range order, represented in fact one of the first indications of 
existence of the pseudogap phase in high-temperature superconductors. 
The DMFT calculation of  $\chi_s$ can be performed 
by evaluating the derivative of the magnetization $m$ with respect to a uniform
magnetic field in the limit of vanishing $h$. In terms of the local 
Green functions
\begin{equation}
\chi_s= \lim_{h \rightarrow 0} \frac{1}{2}T\frac{\sum_{\omega_n} 
\left[G_{\uparrow}(i\omega_n) -G_{\downarrow}(i\omega_n)\right]}{h}.
\end{equation}
This calculations has been performed by varying the temperature in a wide range
($0 <T< 2 D$) for four different values of the pairing interaction 
($U/D= 0.8, 1.8, 2.4$ and $3.6$) and represent an extension of  
the results reported in Ref. [\onlinecite{keller}]. 
The results of our 
calculation are summarized in Fig. \ref{chispin}.

In the weak-coupling side ($U=0.8 D$) we find a conventional metallic 
behavior of  $\chi_s$, which increases monotonically with decreasing 
temperature. The interaction reduces the 
zero-temperature extrapolated value with respect to the non-interacting result
$\chi_s= \rho(0)$. On the opposite side of the phase diagram, in the 
strong-coupling regime ($U= 3.6 D$) 
the standard high-temperature behavior of $\chi_s$ 
extends only down to a certain  temperature $T_M^*$, where a maximum of 
$\chi_s$ is reached. When the temperature is further reduced 
$\chi_s$ starts to decrease, exponentially approaching zero 
in the $T=0$ limit, signaling the opening of a gap in the 
spin-excitation spectrum. A qualitatively similar behavior is found also  
in the intermediate-coupling regime, at least as long as the value of $U$
stays larger than $U_{pairing}$ (e.g., $U=2.4 D$), or, 
in other words, as long as the left line  defining the crossover 
region in Fig. \ref{phd} is not crossed. 
The behavior of  $\chi_s$ becomes richer for $U = 1.8 D < U_{pairing}$. 
At high temperatures $\chi_s$ closely resembles the insulating case, 
displaying 
a clear maximum at a temperature $T_M^*$. By approaching $T=0$,  
$\chi_s$ no longer vanishes, but it rises at small temperatures
displaying a minimum for a temperature lower than $T_M^*$:
a metallic behavior is therefore recovered, 
associated to the narrow resonance at the Fermi level\cite{keller}. Such
a behavior naturally defines a different temperature scale $T_m^*$, 
which is associated to the minimum of $\chi_s$ and represents the lower border
of the pairing zone, or in a sense, of the ``pseudogap'' region. 
Conversely, this low-temperature 
behavior has not been observed to our knowledge in finite-dimensional QMC
simulation \cite{bcsbeqmc,moreo,singerold,singernew,sewer,angleres}.  
In practice, the system displays a pseudogap behavior in the region
between $T_m^*$ and $T_M^*$, whose boundary, labeled as $T_s^*$, is 
represented in Fig. 8. 

We finally mention that the temperature $T_M^*$ for which $\chi_s$ is maximum 
scales with $U$. This finding is in  a qualitative agreement 
with a QMC simulation\cite{singerold,sewer}, 
where the  $T_M^*(U)$ is taken as a definition of the temperature below 
which the  pseudogap appears. 
From a more quantitative point of view, as happens for
$T_c$ and $U_{max}$, the values of $T_M^*(U)$ of the QMC simulations are
lower than our DMFT results (i.e., $T_M^*(U_{max}) \sim 0.15 D$ when  
$d=2$  and $\sim 0.45 D$ for $d=3$, against the DMFT estimate of $\sim 0.7 D$)
. However, also in this case the ratio between  $T_M^*(U_{max})$ and $U_{max}$
has a more universal value around $0.2 \div 0.3$.

\begin{figure}
\begin{center}
\includegraphics[width=8.5cm]{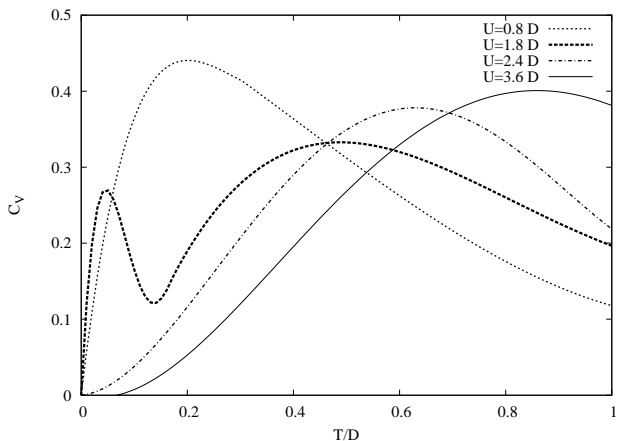}
\caption{Specific heat as a function of temperature for 
$U/D=0.8,1.8,2.4,3.6$. All the $C_V$ lines are obtained by differentiating 
the the internal energy $E_{int}(T)$. The expression of $E_{int}(T)$ is 
computed directly by fitting the DMFT data.}
\label{cv}
\end{center}
\end{figure}

Another relevant quantity is the specific heat 
$C_V=\partial E/\partial T = - T\partial^2 F/\partial T^2$,
that we obtain by differentiating a fit to the DMFT internal energy $E(T)$
 for the same attraction strengths and report in Fig. \ref{cv}. 
Also for this quantity the weak coupling case ($U/D=0.8$) behaves as a regular 
metal, with a linear behavior at small temperatures ($C_V = \gamma T$, 
with $\gamma \propto 1/m^*$, $m^*$ being the effective mass). 
followed by a smooth decrease when the temperature exceeds the
typical electronic energy scale. The same qualitative result is found for 
the noninteracting system, where the low-$T$ slope is smaller since the 
interacting system has a larger effective mass.
In the opposite strong coupling limit we 
observe the typical activated behavior
of gapped systems for small temperatures, with an exponential
dependence of $C_V(T)$ which extends up to a temperature $T^*_{hM}$
large enough to wipe out the effect of the gap. 
It is therefore natural to associate such a temperature to the closure of
the pseudogap.

In the most interesting $U = 1.8D$ 
case, two features are clearly present in the $C_V(T)$ curve. 
The first, low-temperature feature is the evolution of the small-$U$
metallic feature, which acquires a larger slope as $U/D$ is increased due
to the enhancement of the effective mass, and shrinks as a consequence
of the reduced coherence temperature of the metal. 
The second feature is instead the evolution of the large-$U$ insulating
one, and would show an activated behavior partially hidden by the
low-$T$ metallic peak. Thus, the system behaves like a metal in the
small temperature range, while it has a pseudogap for intermediate 
temperature. We estimate the lower boundary of the pseudogap region in
this intermediate coupling regime through the maximum of the low-temperature
feature, which is controlled by the effective coherence scale of the metal.
The upper bound is naturally defined as the temperature in which the
activated behavior disappears. As a result, the specific heat analysis 
determines a pseudogap region with a very similar shape than the one determined
through the spin susceptibility, with a re-entrance of metallic behavior 
in the intermediate coupling regime at low temperatures.

\begin{figure}[htbp]
\begin{center}
\includegraphics[width=8cm]{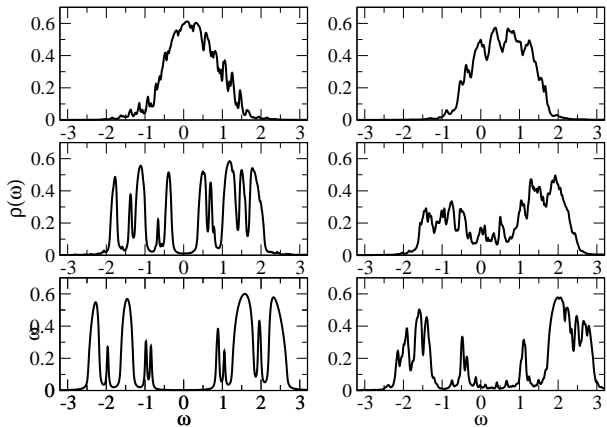} 
\end{center}
\caption{Here are plotted the density of states $\rho(\omega)$ 
for three different values
 of the interaction: $U= 0.8 D, 2.4 D$ and $3.6$ (from the upper to the lower 
row)  both at the low-temperature  ($T=0.1 D$, left panels) 
and at the high-temperature ($T=1 D$, right panels).}
\label{spectral}
\end{figure}

An inspection to the spectral function can strengthen our insight on the 
pseudogap phase. 
In principle  the ED algorithm allows to directly compute finite 
frequency spectral functions $\rho(\omega) = -1/\pi Im G(\omega)$ , 
avoiding the problems and ambiguities
intrinsic with analytic continuation techniques. Unfortunately,
the discretization of the Hilbert space which allows for an ED solution
results in ``spiky'' local spectral functions formed by a collection of 
$\delta$-functions. In this light, we find it useful to compute
 $\rho(\omega)$ by  analytically continuing Eq.(\ref{sigma}), 
analytically computing the local retarded Green function
$G_{loc}(\omega)= -1/\pi\int d\epsilon D(\epsilon) 
(\omega - \epsilon +\mu - \Sigma_{ret}(\omega))^{-1}$. 
This procedure provides more ``realistic''
descriptions of both the non-interacting DOS and of the strong-coupling
pairing phase.  However, even though the spectral functions are  
smoothed by this procedure, we can only extract informations about the
gross features of the spectra as, e.g., the amplitude of the gap $\Delta(T)$.
Keeping these limitations in mind, some results for $\rho(\omega)$ are plotted
in Fig. \ref{spectral}: for the weak- ($U=0.8 D$), 
the intermediate-  ($U=2.4 D$) and strong-coupling ($U=3.6 D$) case 
a low and high-temperature set of data are shown. 
Apart from the obvious appearing and enlarging of a gap in  $\rho(\omega)$
 with 
increasing $U$, which is evident in the low-temperature data, it should be
noticed that both in the intermediate and the strong-coupling regime there is
apparently no tendency to a 'closure' of such a gap when the temperature is 
raised. Indeed, as it is shown in the second and the third row of  Fig. 
\ref{spectral}, the gap starts to fill at some temperature
($T \sim 0.45 D$ for $U =2.4 D$ and $T \sim 1.5\div 2.0 D$ 
for $U=3.6 D$), but for these values of $U$ 
much of the spectral weight remains in the high-energy Hubbard bands,
and the gapped structure  does not completely 
vanish up to the highest temperature reached in our calculation 
($T \simeq 2D$). 
On the other hand, QMC results in $d=2$\cite{singernew} 
obtained through maximum entropy show a closure of the gap in $\rho(\omega)$ 
at  a temperature lower that our threshold.    
Further investigation is needed to understand whether the discrepancy is due
to a different behavior between $d=2$ and the infinite dimensionality limit,
or it is determined by the technical difficulties involved in the calculation
of real frequency spectra in both approaches.
The persistence of the gap structure at high temperature
that we find in DMFT is also obtained within a perturbative analysis of
superconducting fluctuations at strong coupling in $d=2$\cite{pera}.

\begin{figure}
\begin{center}
\includegraphics[width=8.5cm]{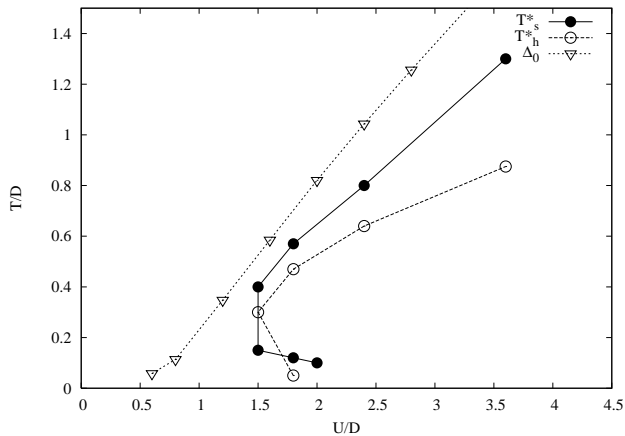}
\caption{Different estimates of the pseudogap temperature. The filled 
circles indicates $T_s^*(U)$, i.e., the temperatures of both the 
maxima and the minima of $\chi_s(T)$, while the empty circle marked $T_h(U)$
that is the temperature associated with the maxima and
the minima of $C_V(T)$. The regions on the left of these two lines can be 
interpreted as the zone of ``pseudogap'' behavior for the spin and the 
specific heat respectively. These lines are then compared with the behavior
of the anomalous part of the self-energy at zero temperature 
($\Delta_0$, empty triangles).}
\label{tstar}
\end{center}
\end{figure}

In Fig. \ref{tstar} we compare our estimates of  the pseudogap 
temperature obtained through different physical quantities.
We draw the borders of the pseudogap region as determined from
the spin susceptibility ($T^*_s(U)$) and the specific-heat behavior ($T_h^*(U)$
and the value of the superconducting gap $\Delta_0$ at zero temperature:
The upper borders of the spin and the specific-heat ``pseudogap'' region scale
roughly with $U$, as $\Delta_0$ does, so that both  $T_h^*(U)$ and $T_s^*(U)$
are proportional to $\Delta_0$, as the experimentally determined pseudogap.
At low temperature, the pseudogap region boundary as extracted from 
thermodynamic response functions displays a clear re-entrance, which 
can be associated with the onset of the low-temperature quasi-particle 
peak. We also notice that the low-temperature curve qualitatively follows the
behavior of the $U_{c2}(T)$ line. As mentioned above, the slope of
$U_{c2}(T)$ is easily interpreted in terms
of entropy balance between the two phases, which favors the preformed pairs
phase. 

Our phase diagram also represents a warning regarding attempts to extrapolate
the low-temperature behavior from the high-temperature data in order to 
compare with finite-dimensional QMC calculations. 
If one, as, e.g., in Ref. \onlinecite{sewer},  extrapolated the 
high-temperature behavior down to $T=0$ in order to estimate the 
metal-insulator point, would have obtained an estimate
of $U^*$ significantly lower than the real $U_{c2}$.
This finding emphasize how the high-temperature properties of 
the attractive Hubbard model are only weakly dependent on dimensionality, 
as indicated by the similarity between DMFT and finite-dimension QMC, 
while the low-temperature behavior may well be dependent on the dimensionality,
as well as on the details of the bandstructure of the underlying lattice.
 
\section{Conclusions}
In this paper we have investigated the finite temperature aspects
of pairing and superconductivity in the attractive Hubbard model
by means of DMFT,  considering both normal and superconducting solutions.

In the normal phase we have identified two families of solutions, a
Fermi-liquid metallic phase and a preformed-pair phase with 
insulating character. 
The latter phase is formed by local pairs without phase coherence. 
A finite region of the coupling-temperature
phase diagram is characterized by the simultaneous presence of both 
solutions. In the low temperature regime a first-order transition occurs 
within this region when 
the free energies of the two solutions cross, and the region closes
at a certain temperature ($T_{pairing} =0.03 D$) in a critical point. 
Interestingly, some trace
of the two solutions survives even for temperature larger than the
critical temperature, and two crossover lines can be defined
separating a normal metal, a sort of semiconductor in which the gap is 
closed by temperature, and the preformed-pair phase with
a well defined gap.

When superconductivity is allowed, the superconducting solution is stable
for all values of the attraction and the critical temperature is 
always larger than the pairing transition temperature in the normal phase.
In the superconducting state, we find an evolution from a weak-coupling
BCS-like behavior, with exponentially small $T_c$ 
from a normal metal to the superconductor, and a strong-coupling regime
in which superconductivity is associated to the onset of the phase-coherence
among the  preformed pairs that occurs at $T_c \propto t^2/U$. 
The highest $T_c$ is obtained in the intermediate region between this
two limiting cases, namely for $U \simeq 2.1 D$, which is extremely 
close to the zero-temperature critical point of the normal phase.

The presence of the pairing transition affects the normal phase above $T_c$ 
also when superconductivity establishes. 
In particular, one could be tempted to identify the phase of preformed 
pairs obtained at strong coupling with
the pseudogap behavior observed in cuprates.
In order to test the adequacy of such an identification, we computed 
different observables, whose anomalies can identify the appearance of the
pseudogap, like the spin susceptibility, the specific heat and the
single particle spectral functions. 
In the intermediate region of coupling, where the pairing transition 
occurs and the superconducting critical temperature reaches its
maximum, the pseudogap region presents a re-entrance at low
temperatures associated with a small coherent peak in the spectral 
function. At temperatures smaller than this coherence temperature
the system behaves like a normal metal with renormalized effective
mass. 
On the other hand, the high-temperature boundary of the pseudogap
region scales with $U$ regardless the criterion we use to estimate it.
The estimate of the pseudogap temperature from specific heat and
spin susceptibility both scale with the zero-temperature gap, as
in the cuprates.

The most striking difference between our pseudogap 
phase-diagram and the experiments in the cuprates is 
that the pseudogap phase in the attractive Hubbard model is much larger 
than the experimental one, as it is measured by the large 
value of $T^*_{s,h}/T_c \simeq 5$ at the optimal value of the attraction.
The experimental $T^*$ around optimal doping is instead very close to 
$T_c$, and, according to some authors, the pseudogap line tends to zero
at optimal doping.
Moreover, the pseudogap temperature observed in the cuprates is 
definitely much smaller than the one found within our DMFT of the 
attractive Hubbard model.
This inadequacy of the attractive Hubbard model in describing 
some features of the pseudogap phase descend from the 
above mentioned strong simplifications of the model 
(neglect of retardation effects, Coulomb repulsion and d-wave symmetry of 
the gap) and of our DMFT treatment which is exact only in the infinite
dimensionality limit.
One could be tempted in maintaining an attractive Hubbard model description
for the quasiparticles alone, but it is important to point out
that this interpretation can not be pushed too far. As an example, 
it is clear that such a description would fail for 
temperatures larger than the quasiparticle renormalized bandwidth.

A better description of the pseudogap phase would require models 
in which both an attraction and a repulsion are present.
This is for instance the case of the models introduced in Refs. 
\onlinecite{science,exe}, where the superconducting phenomenon
only involves heavy quasiparticles which experience an unscreened
attraction and a richer behavior of the pseudogap 
(which in this case closes around optimal doping) is found.

\section{acknowledgments}
This work is also supported by MIUR Cofin 2003.
We acknowledge useful discussions with S. Ciuchi, M. Grilli, 
and G. Sangiovanni.

\end{document}